\documentclass{article}

\usepackage{amsmath}
\usepackage{epsfig}
\usepackage{hyperref}
\usepackage{placeins}
\usepackage{fullpage}

\begin{document}
\title{Augmenting Recommendation Systems Using a Model
of Semantically-related Terms Extracted from User Behavior}

\author{
Khalifeh AlJadda*, Mohammed Korayem**, Camilo Ortiz, \\ Chris Russell, David Bernal, Lamar Payson,\\ Scott Brown, and Trey Grainger\\\\
       {\normalsize Search Relevancy and Recommendations, CareerBuilder}\\
       {\normalsize *Department of Computer Science,  University of Georgia, GA}\\
	   {\normalsize **School of Informatics \& Computing, Indiana University, IN} \\\\       	       
       {\small khalifeh.aljadda, mohammed.korayem, camilo.ortiz@careerbuilder.com}\\
       {\small chris.russell, david.bernal, lamar.payson@careerbuilder.com}\\
       {\small scott.brown, trey.grainger@careerbuilder.com}    
}

\date{}


\maketitle
\begin{abstract}
Common difficulties like the cold-start problem and a lack of sufficient information about users due to their limited interactions have been major challenges for most recommender systems (RS).
To overcome these challenges and many similar ones that result in low
accuracy (precision and recall) recommendations, we propose a novel
system that extracts semantically-related search keywords based
on the aggregate behavioral data of many users. These semantically-related search keywords
can be used to substantially increase the amount of knowledge about a specific user's interests based upon even a few searches and thus improve the accuracy of the RS. The proposed system is capable
of mining aggregate user search logs to discover semantic relationships between
key phrases in a manner that is language agnostic, human understandable,
and virtually noise-free. These semantically related keywords are
obtained by looking at the links between queries of similar users
which, we believe, represent a largely untapped source for discovering
latent semantic relationships between search terms.
\end{abstract}



\section{Introduction}

Recommender systems are widely used these days in many industries
like e-commerce, video streaming, and job portals. Recommender systems
automate the process of discovering the interests of a user and subsequently
what is relevant to his/her needs \cite{konstan2004introduction,park2009pairwise,sarwar2001item}.

Many companies like Netflix%
\footnote{http://www.netflix.com%
}, Amazon%
\footnote{http://www.amazon.com%
}, CareerBuilder%
\footnote{http://www.careerbuilder.com%
}, etc. depend on recommender systems (RS) to help drive their
revenue. For example, Netflix, a movie rental and video streaming web
site, offered a prize (known as the Netflix prize) of 1 million dollars
in 2006 for any recommendation algorithm that could beat their RS,
named Cinematch\cite{bennett2007netflix}. Netflix, like many other
websites, depends heavily on recommendations in order to keep their
customers interested in their service. One of the major challenges
for any RS is the cold-start problem~\cite{park2009pairwise}, which
occurs when there is a lack of information linking new users or items
such that the RS is unable to determine how those users or items
are related and is therefore unable to provide useful recommendations.
One possible way to solve this is to perform classification or matching
based upon the description (i.e. keywords in the text) of the items
which could be recommended, but this approach tends to not perform
as well as a collaborative filtering approach which links related
items together based upon the collective intelligence of many users.

Instead of directly linking users and items, we propose a system that
utilizes the wisdom of the crowd. This system incorporates a language-agnostic
technique for discovering the intent of user searches by revealing
the latent semantic relationships between the terms and phrases within
keyword queries. Our hope is to express these relationships in common
human language so that we can automatically augment recommendations
when the only data available from a user is few search keywords extracted
from the searches that he conducted. Mining the search history of
millions of users allows us to discover the relationship between search
terms and the most common meaning of each term. Once we know the semantic
relationships between terms according to these users, we can then
use those relationships to enhance the recommendation features in
order to more accurately express the interest of each user. 

Our use case for the proposed technique is for the recommendation system CareerBuilder, which operates
the largest job board in the U.S. and has an extensive and growing
global presence, with millions of job postings, more than 60 million
actively-searchable resumes, over one billion searchable documents,
and more than a million searches per hour. Our solution to this challenge
makes use of the wisdom of the crowd to discover domain-specific relationships.
Using the query logs of more than a billion user searches,
we can discover the family of related keyword phrases for any particular
term or phrase for which a search has been conducted. We are not attempting
to fit these terms into an artificial taxonomy, instead we are discovering
the existing relationships between terms according to our users.

\section{Search Log Analyzer}

Our proposed system is applicable for websites and services that receive
searches from massive numbers of users (i.e. millions), such as is the case with CareerBuilder.com.
We propose a \emph{search log analyzer} (SLA) that aims to discover
latent semantic relationships among the users' search terms in order
to build a semantic dictionary that can more expressively interpret
a user's query intent which in turn provide more relevant results
in both the search engine and RS. One important feature of the proposed
SLA is that it is language agnostic, as there is no dependency on
any language-specific natural language processing (NLP) techniques.
This makes the system applicable on any website or system that receives
a large number of searches, regardless of the list of languages the
system supports.

\begin{figure}
\centering \epsfig{file=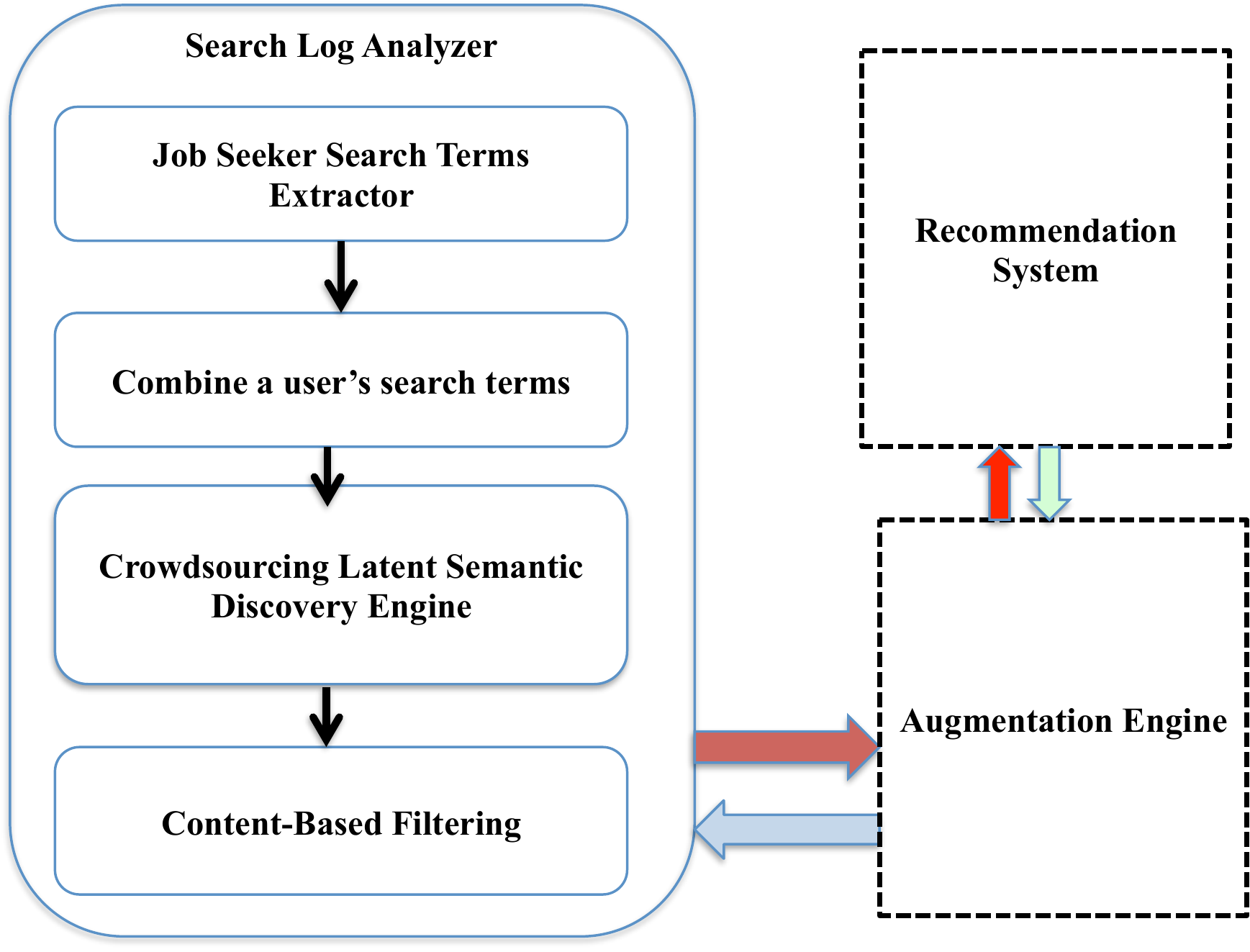,scale=0.4} \protect\caption{\label{fig:sysArch}System Architecture}
\end{figure}

\begin{table}
\centering \protect\caption{Input data to PGMHD}

\label{input_pgmhd} %
\begin{tabular}{l|p{2.5cm}|p{3.5cm}}
\hline 
UserID  & Classification  & Search Terms \tabularnewline
\hline 
\hline 
user1  & Java Developer  & Java, Java Developer, C, Software Engineer\tabularnewline
\hline 
user2  & Nurse  & RN, Rigistered Nurse, Health Care\tabularnewline
\hline 
user3  & .NET Developer  & C\#, ASP, VB, Software Engineer, SE\tabularnewline
\hline 
user4  & Java Developer  & Java, JEE, Struts, Software Engineer, SE\tabularnewline
\hline 
user5  & Health Care  & Health Care Rep, HealthCare\tabularnewline
\hline 
\end{tabular}
\end{table}

\begin{figure}
\centering \epsfig{file=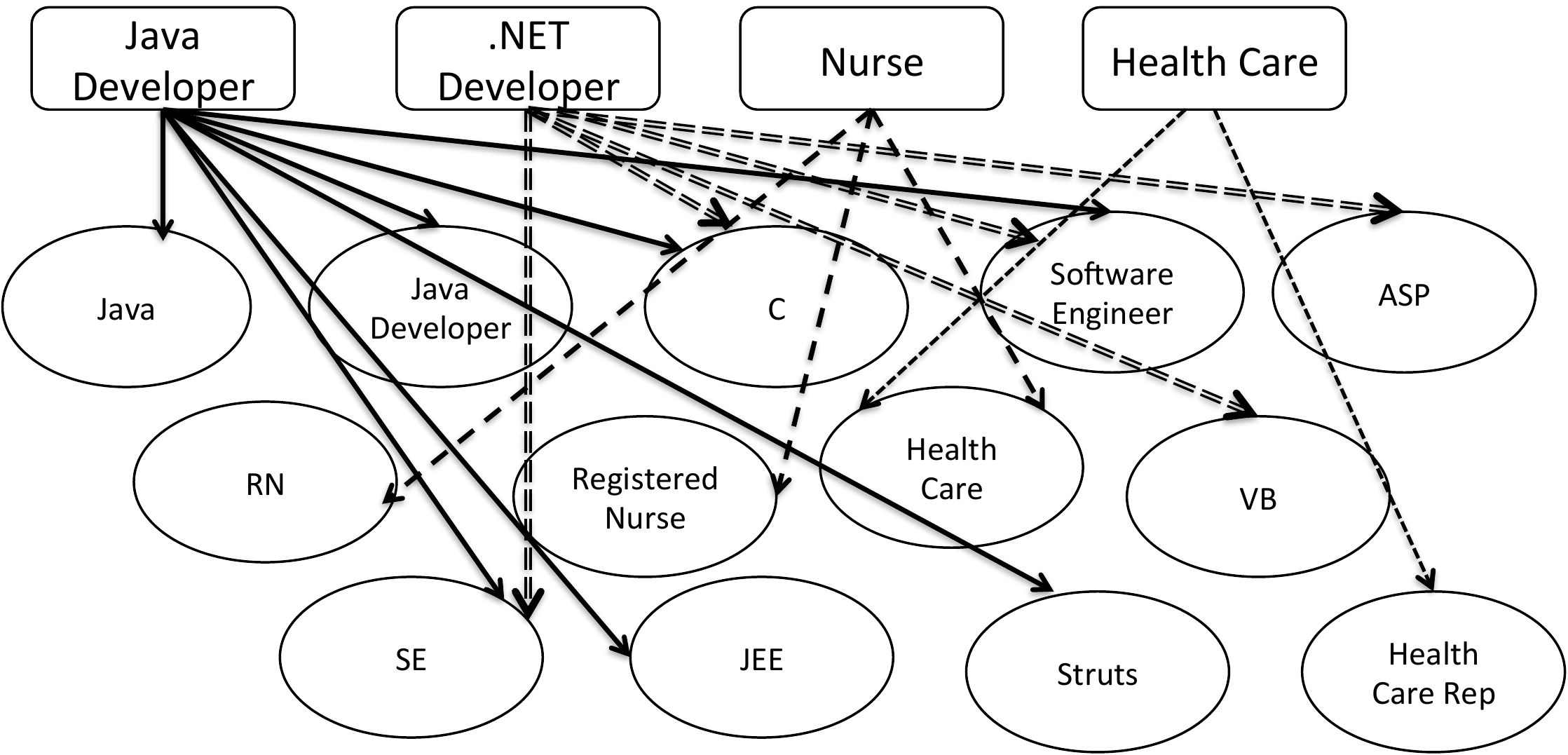,scale=0.4} \protect\caption{\label{pgmhdJobs}PGMHD representing the search log data}
\end{figure}

\begin{figure*}[th]
\centering \epsfig{file=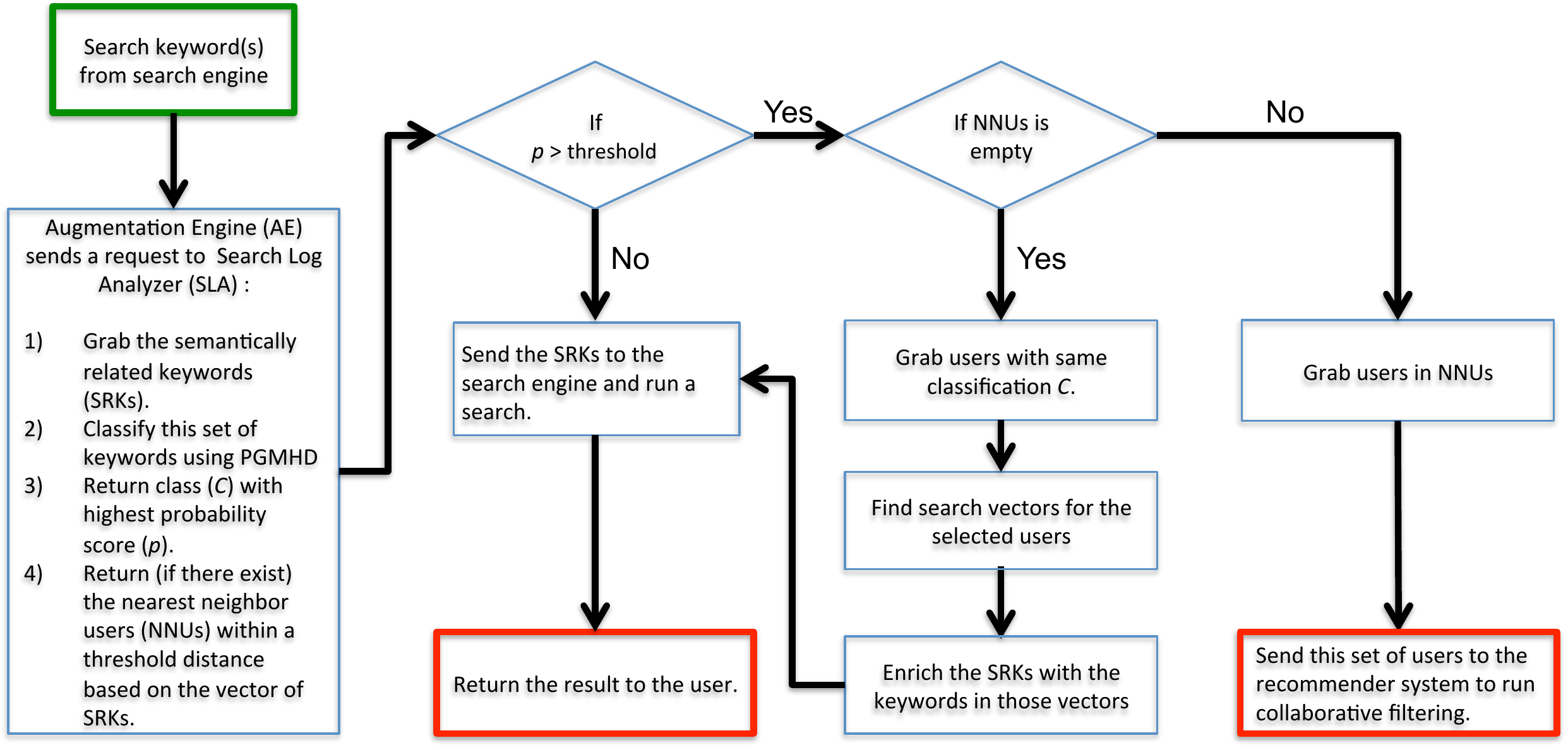,scale=0.6} \protect\caption{\label{AugArch} Augmentation Engine Architecture}
\end{figure*}

\subsection{Crowdsourced Latent Semantic Discovery Engine}

The most significant component in the proposed SLA is the crowdsourced
latent semantic discovery engine. This engine is designed in a way
that enables the utilization of crowdsourced wisdom. The model we
used to represent the data is a probabilistic graphical model for
massive hierarchical data (PGMHD) \cite{DBLP:journals/corr/AlJaddaKOGMY14}, which we designed and implemented
to obtain a scalable variant of a Bayesian Network that is suitable
for this kind of massive data. In order to represent search log data
in this model, pre-processing phases are required as shown in Figure~\ref{fig:sysArch}.
The pre-processing phases include:
\begin{enumerate}
\item classify the users.
\item grab the query strings of their searches.
\item extract the search terms of those query strings.
\item aggregate the search terms of each user.
\item combine the user's classification with his search terms in one table to be used as input table to build the PGMHD.
\end{enumerate} 
Table~\ref{input_pgmhd} shows the processed data to be represented
using PGMHD and Figure~\ref{pgmhdJobs} shows an example of PGMHD
representing search log data. The root nodes in this model represent
the classification of the users who conducted searches. Additionally,
the second level nodes represent the keywords used in the searches
conducted by the users. An edge between a root node (user's classification)
and a child node (search keyword) represents the usage of a search
keyword by the users from a classification. 

To obtain the estimates in our probabilistic model, we store the number
of searches $f_{ck}$ for a keyword $k$ by users of class $c$ at
every edge that connects $c\rightarrow k$. This way, we can naturally
estimate the joint probabilities 
\[
P(k,c)=\frac{f_{ck}}{\underset{ij}{\sum}f_{ij}},
\]
and similarly, the conditional probabilities 
\[
P(k|c)=\frac{f_{ck}}{\underset{j}{\sum}f_{cj}},\quad P(c|k)=\frac{f_{ck}}{\underset{i}{\sum}f_{ik}},
\]
required by the PGMHD. The SLA is implemented using the following
technologies: HDFS~\cite{shvachko2010hadoop}, Map/Reduce jobs~\cite{dean2008mapreduce},
Hive~\cite{thusoo2009hive}, and Solr%
\footnote{http://lucene.apache.org/solr%
}.

\begin{table*}[th]
\caption{Sample Results using Solr for content-based filtering (CBF)}

\label{solar}
\begin{tabular}{ l | p{4.2cm} | p{4.2cm} | p{4.2cm} }
\hline
\textbf{Term} & \textbf{Before Solr} & \textbf{Removed Terms (Outlier)} & \textbf{Final List}  \\
\hline

data scientist &
machine learning, data analyst, data mining, analytics, big data, statistics, ... &
data analyst, data mining, analytics, statistics &
machine learning, big data \\

\hline

cashier&
retail, retail cashier, customer service, cashiers , receptionist , cashier jobs, teller, ... &
receptionist, teller &
retail, retail cashier, customer service, cashiers, cashier jobs \\

\hline
collections Rep &
collector, collections specialist, call center rep, credit and collections, ... &
call center rep &
collector, collections specialist, credit and collections\\

\hline
repair tech &
repair technician, maintenance technician, call center &
maintenance technician, call center &
repair technician\\

\hline
front end development &
web developer, productivity, Monster  &
productivity, monster  &
web developer\\

\hline
\end{tabular}
\end{table*}

\subsection{Content-based Filtering}

In our content-based post-filtering phase, we apply the search engine~\cite{solarinaction}
itself as part of the SLA for post-filtering. Since our ultimate goal
is to find the most related search terms that have actual semantic
similarity, we examine all of the relationships for each term. For
each relationship we run a query using the original term and another
query using the related term. We compare the two resulting document
sets and look for intersection. If there are too few intersections
we consider the relationship to be invalid and remove it. Table \ref{solar} shows how the content-based filtering improves the accuracy of the discovered semantic relationships between search terms.

\section{Augmentation Engine}

There are several approaches to follow in order to improve the accuracy
of the RS using the proposed SLA. In our design, the \emph{augmentation
engine} (AE) decides which combination of approaches to follow in order to improve
the recommendations generated by the RS. Our implementation of the
AE focuses on deciding which combination of the following three augmentations
(to the RS) should be used: 
\begin{enumerate}
\item a \emph{search query augmentation},
\item a \emph{user classification augmentation}, or
\item a \emph{nearest neighbor augmentation}. 

\end{enumerate}
These three augmentation approaches try to tackle the special ``cold-start''
cases when the only information you have about a user are the search
queries that he/she performs through the search engine. The next three
subsections describe in detail each of the augmentation approaches
mentioned above.

\subsection{Search Query Augmentation}

In the case when the RS is not able to recommend items to a user
(possibly due to the ``cold-start'' problem), it is possible to use the search
engine to assist the RS based on the user's search keywords. However, it might be the case that the search
keywords from the user are not enough for the search engine to retrieve
a significant amount of items to feed to the RS. In such a case, the semantically-related
keywords represent a crucial source of information to augment the
search query. The \emph{search query augmentation} that we\emph{ }propose
uses the semantically-related keywords obtained from the SLA to enhance
the search queries that will be used on the search engine that feeds
the RS.

\subsection{User Classification Augmentation}

In the previous subsection, we presented a simple augmentation approach
that only enriches the search query for the search engine with the
semantically related search keywords, ignoring any information that
can be obtained about the users who share similar searching behavior.
To go beyond that basic search query augmentation, we propose to
utilize the search keywords to classify the users with the PGMHD model
\cite{DBLP:journals/corr/AlJaddaKOGMY14} used within the SLA. Given
a set of search keywords $K$ and a set $C$ of pre-defined classifications
of the users, PGMHD calculates the probability scores $p_{c}$ for
every class of user $c\in C$. The \emph{user classification augmentation
}that we\emph{ }propose can be combined with the \emph{search query augmentation
} described in the last subsection to refine the decision space in order to return more precise results which fall within the same classification as the user.

\subsection{Nearest Neighbor Augmentation}

We can further improve the precision of the recommendations obtained
from the user classification augmentation presented in the previous
subsection by considering a specific subset of the users from the
class $c_{\max}$ based on a $k$--nearest neighbor approach. More
specifically, by representing each user as vector of its search keywords,
we can define a distance (e.g., Hamming, Euclidean, etc.) between
a given user and other users, so as to discover the most similar/nearest
ones. The \emph{nearest neighbor augmentation} that we\emph{ }propose
enables the RS to use a collaborative filtering (CF) approach in a
more precise way by using the most similar users from the class $c_{\max}$.

\subsection{Augmentation Engine Architecture}

Let $K$ be the set of most relevant search keywords from the user
for which the RS will provide recommendations, and $C$ be the set
of pre-defined classifications of the users. Then, the proposed AE
operates as follows (see Figure \ref{AugArch}). First, based on the
set of search keyword(s) $K$, the AE sends a request to the SLA to
obtain: i) a set $R$ of semantically related keywords; ii) the probability
scores $p_{c}$ obtained from a PGMHD classification of the keywords,
for every class of user $c\in C$; and iii) the set of nearest neighbor
users (NNUs) from the class $c_{\max}$ within a threshold distance.
If the maximum probability score is larger than a given threshold
$\alpha>0$, i.e. 
\begin{equation}
\max_{c\in C}p_{c}=p_{c_{\max}}>\alpha,\label{eq:prob-thresh}
\end{equation}
and the set of NNUs is not empty, a \emph{nearest neighbor augmentation}
is chosen. If \eqref{eq:prob-thresh} is satisfied but the set of
NNUs is empty, a \emph{user classification augmentation} is applied (to refine the decision space), and a \emph{search query augmentation} is applied within the decision space corresponding with the user's classification.
Otherwise, if \eqref{eq:prob-thresh} is not satisfied, a \emph{search
query augmentation} is chosen without restricting the decision space to a specific classification corresponding to the user.

\section{Experiment and Results}

We have implemented the Search Log Analyzer (SLA) using Hadoop Map/Reduce
framework. The SLA was applied to analyze 1.6 billion search logs
obtained from CareerBuilder.com. After applying the content-based
filtering, the discovered related search terms had an error rate of
at most 2\%. Table~\ref{pgmhdResults} shows sample results of the
SLA. Currently, we are implementing tests on the RS that incorporate
the proposed AE to evaluate the impact of the proposed system at CareerBuilder.com.

\begin{table}[h!]
\begin{center}
\protect\caption{Results of SLA}

\label{pgmhdResults} %
\begin{tabular}{l|p{5cm}}
\hline 
Term  & Related Terms\tabularnewline
\hline 
hadoop  & big data, hadoop developer, OBIEE, Java, Python\tabularnewline
\hline 
registered nurse  & rn registered nurse, rn, registered nurse manager, nurse, nursing,
director of nursing\tabularnewline
\hline 
data mining  & machine learning, data scientist, analytics, business intellegence,
statistical analyst\tabularnewline
\hline 
Solr  & lucene, hadoop, java\tabularnewline
\hline 
Software Engineer  & software developer, programmer, .net developer, web developer, software \tabularnewline
\hline 
big data  & nosql, data science, machine learning, hadoop, teradata \tabularnewline
\hline 
Realtor  & realtor assistant, real estate, real estate sales, sales, real estate
agent \tabularnewline
\hline 
Data Scientist  & machine learning, data analyst, data mining, analytics, big data\tabularnewline
\hline 
Plumbing  & plumber, plumbing apprentice, plumbing maintenance, plumbing sales,
maintenance\tabularnewline
\hline 
Agile  & scrum, project manager, agile coach, pmiacp, scrum master\tabularnewline
\hline 
\end{tabular}

\end{center}
\end{table}


\section{Conclusions}

In this paper we propose a novel system to augment recommendations
generated by a RS when the only available information about a user
for recommendations is a small set of searched keywords. The proposed
system relies on a model that discovers the semantic relationships
between search keywords using aggregate behavioral data from millions
of users. Moreover, the proposed system is language-agnostic and could
be integrated with most modern recommendation engines. \\ \\


\section{Acknowledgments}

We would like to greatly thank the Big Data and Data Science teams
at CareerBuilder for their support and feedback during implementation
of the Search Log Analyzer. Also, many thanks to the Search Development
Group at CareerBuilder for their help with integrating the proposed
system within the recommendation engine. 

\bibliographystyle{abbrv}
\bibliography{sigproc}

\begin{thebibliography}{1}

\bibitem{DBLP:journals/corr/AlJaddaKOGMY14}
K.~AlJadda, M.~Korayem, C.~Ortiz, T.~Grainger, J.~A. Miller, and W.~S. York.
\newblock {PGMHD}: A scalable probabilistic graphical model for massive
  hierarchical data problems.
\newblock {\em CoRR}, abs/1407.5656, 2014.

\bibitem{bennett2007netflix}
J.~Bennett and S.~Lanning.
\newblock The netflix prize.
\newblock In {\em Proceedings of KDD cup and workshop}, volume 2007, page~35,
  2007.

\bibitem{dean2008mapreduce}
J.~Dean and S.~Ghemawat.
\newblock Mapreduce: simplified data processing on large clusters.
\newblock {\em Communications of the ACM}, 51(1):107--113, 2008.

\bibitem{solarinaction}
T.~Grainger and T.~Potter.
\newblock {\em Solr in Action}.
\newblock Manning Publications Co., 2014.

\bibitem{konstan2004introduction}
J.~A. Konstan.
\newblock Introduction to recommender systems: Algorithms and evaluation.
\newblock {\em ACM Transactions on Information Systems (TOIS)}, 22(1):1--4,
  2004.

\bibitem{park2009pairwise}
S.-T. Park and W.~Chu.
\newblock Pairwise preference regression for cold-start recommendation.
\newblock In {\em Proceedings of the third ACM conference on Recommender
  systems}, pages 21--28. ACM, 2009.

\bibitem{sarwar2001item}
B.~Sarwar, G.~Karypis, J.~Konstan, and J.~Riedl.
\newblock Item-based collaborative filtering recommendation algorithms.
\newblock In {\em Proceedings of the 10th international conference on World
  Wide Web}, pages 285--295. ACM, 2001.

\bibitem{shvachko2010hadoop}
K.~Shvachko, H.~Kuang, S.~Radia, and R.~Chansler.
\newblock The hadoop distributed file system.
\newblock In {\em Mass Storage Systems and Technologies (MSST), 2010 IEEE 26th
  Symposium on}, pages 1--10. IEEE, 2010.

\bibitem{thusoo2009hive}
A.~Thusoo, J.~S. Sarma, N.~Jain, Z.~Shao, P.~Chakka, S.~Anthony, H.~Liu,
  P.~Wyckoff, and R.~Murthy.
\newblock Hive: a warehousing solution over a map-reduce framework.
\newblock {\em Proceedings of the VLDB Endowment}, 2(2):1626--1629, 2009.

\end{thebibliography}

\end{document}